\documentclass[aps,pre,floatfix,twocolumn,superscriptaddress]{revtex4-1}
\usepackage{color}
\usepackage{graphicx}
\usepackage{subfigure,amsmath, amsthm, amssymb}
\usepackage[colorlinks,citecolor=red]{hyperref}

\newcommand{\be}{\begin{equation}}
\newcommand{\ee}{\end{equation}}
\newcommand{\bea}{\begin{eqnarray}}
\newcommand{\eea}{\end{eqnarray}}

\begin{document}
\title{Phases of the hard-plate lattice gas on a three-dimensional cubic lattice}
\author{Dipanjan Mandal}
\email{dipanjan.mandal@warwick.ac.uk}
\affiliation{Department of Physics, University of Warwick, Coventry CV4 7AL, United Kingdom}
\author{Geet Rakala}
\email{geet.rakala@oist.jp}
\affiliation{Okinawa Institute of Scince and Technology, 1919-1 Tancha, Onna-son, Kunigami-gun, Okinawa-ken, Japan}
\author{Kedar Damle}
\email{kedar@theory.tifr.res.in}
\affiliation{Department of Theoretical Physics, Tata Institute of Fundamental Research, Mumbai 400 005, India}
\author{Deepak Dhar}
\email{deepak@iiserpune.ac.in}
\affiliation{Indian Institute of Science Education and Research, Dr. Homi Bhabha Road, Pashan,  Pune 411008, India}
\author{R. Rajesh} 
\email{rrajesh@imsc.res.in}
\affiliation{The Institute of Mathematical Sciences, C.I.T. Campus, Taramani, Chennai 600113, India}
\affiliation{Homi Bhabha National Institute, Training School Complex, Anushakti Nagar, Mumbai 400094, India}

\date{\today}
\begin{abstract}
We study the phase diagram of a system of $2\times 2\times 1$ hard plates on the three dimensional cubic lattice, {\em i.e.} a lattice gas of plates that each cover an elementary plaquette of the cubic lattice and occupy its four vertices, with the constraint that no two plates occupy the same site of the cubic lattice. We focus on the isotropic system, with equal fugacities for the three orientations of plates.
We show, using grand canonical Monte Carlo simulations, that the system undergoes two density-driven phase transitions with increasing density of plates: the first from a disordered fluid to a layered phase,  and the second from the layered phase to a sublattice-ordered phase. 
In the layered phase, the system breaks up into disjoint slabs of thickness two along one spontaneously chosen cartesian direction. Plates with normals perpendicular to this layering direction are preferentially contained entirely within these slabs,  while plates straddling two successive slabs have a lower density. This corresponds to a two-fold symmetry breaking of translation symmetry along one spontaneously chosen cartesian direction, leading to ``occupied slabs'' stacked along the layering direction with a separation of one lattice spacing.  Additionally the symmetry between the three types of plates is spontaneously broken, as plates with normal along the layering direction have a lower density than the other two types of plates. Intriguingly, the occupied slabs exhibit two-dimensional power-law columnar order even in the presence of a nonzero density of vacancies. In contrast, inter-slab correlations of the two-dimensional columnar order parameter decay exponentially with the separation between the slabs. 
In the sublattice-ordered phase, there is two-fold ($Z_2$) breaking of lattice translation symmetry along all three cartesian directions.   
We present numerical evidence that  the disordered to layered transition is continuous and consistent with the three-dimensional $O(3)$ universality class, while the layered to sublattice transition is first-order in nature.

\end{abstract}
\maketitle

\section{Introduction}

Systems of particles interacting through only excluded volume interaction may exist in different phases depending on the shape
and density of the particles. These find a variety of applications, including in self-assembly~\cite{van2006colloids,2012-science-deg-predictive,2017-mpolpp-nc-observation}, efficient drug delivery~\cite{2007-jys-jcr-particle,2017-j-ijp-effect}, design of novel materials with specific optical and chemical properties~\cite{2001-vbsn-nature-onchip,2011-fpmnsocfd-acsn-assembly,2013-skrckpy-nature-shaping}, design of
molecular logic gates~\cite{2011-acsn-smrmsahecj-manipulating,2011-prb-smsacrmehj-demonstration, 2013-acsn-gkkszsmej-contacting}, and
adsorption of gas on metallic surfaces~\cite{1985-prb-twpbe-two,2000-ssr-psb-phase,1991-jcp-dmr-model}. More generally, they are of interest as simple models of fluids~\cite{solana2013perturbation} as well as being the simplest systems to study critical behavior.
Many shapes have been studied in the literature. These include different types of polyhedra~\cite{2012-science-deg-predictive}, colloidal superballs~\cite{2017-mpolpp-nc-observation}, and rods~\cite{1992-vl-rpp-phase}.

Parallel to the study of models in the continuum, models of hard-core particles on lattices, known as hard core lattice gases (HCLGs)   have also been studied.
In literature, many different geometrical shapes have been studied on two dimensional lattices, including triangles~\cite{1999-vn-prl-triangular}, 
squares~\cite{1967-bn-jcp-phase,1966-bn-prl-phase,1966-rc-jcp-phase,2012-rd-pre-high,2016-ndr-epl-stability,2017-mnr-jsm-estimating},
dimers~\cite{1961-k-physica-statistics,1961-tf-pm-dimer,2003-hkms-prl-coulomb,2017-naq-arxiv-polyomino}, Y-shaped particles~\cite{2018-mnr-pre-phase},
mixture of squares and dimers~\cite{2015-rdd-prl-columnar,2017-mr-pre-columnar},
rods~\cite{2007-gd-epl-on,2013-krds-pre-nematic}, 
rectangles~\cite{2014-kr-pre-phase,2015-kr-pre-asymptotic,2015-nkr-jsp-high,2017-gvgmv-jcp-ordering},
discretised discs or the k-NN model~\cite{2007-fal-jcp-monte,2014-nr-pre-multiple,darjani2019liquid,thewes2020phase,jaleel2021hard},
and hexagons~\cite{1980-b-jpa-exact}, the last being the only exactly solvable model. 
A variety of different ordered phases may be observed, including crystalline, columnar or striped, nematic, and power-law correlated phases. Though many examples exist, it is not clear {\em a priori} which phases are realized and in what order (as a function of increasing density) for a given shape.

Comparatively less is known about HCLG models in three dimensions. A detailed phase diagram that encompasses all densities is known for 
only rods of shape $k\times 1 \times1$~\cite{2017-vdr-jsm-different,2017-gkao-pre-isotropic} or $2\times2\times2$ hard cubes~\cite{vigneshwar2019phase}. The numerical study of HCLG models  is constrained by difficulties of  equilibrating the system at densities close to the maximal possible density, as the system gets stuck in very long-lived metastable systems. These difficulties are substantially reduced by using Monte Carlo algorithms that include cluster moves~\cite{2014-kr-pre-phase,2015-kr-pre-asymptotic,2015-rdd-prl-columnar}, which significantly decrease the autocorrelation times. 

Systems of plates or board-like particles in the continuum have also been studied numerically~\cite{2017-cdmp-sm-phase,2011-mvv-pccp-biaxial,2018-dtdrd-prl-hard}. 
The phase diagram in the continuum is very rich, showing multiple transitions with increasing particle densities, and varying aspect ratios. A variety of different phases arise, including smectic, biaxial smectic, uniaxial and biaxial nematic, and columnar with alignment along the long or short axis. If the orientations of the plates are restricted to orthogonal cartesian directions, then it is possible to obtain some rigorous results regarding the nature of the phases, in particular for
a system of hard parallelepipeds of size $1\times k^\alpha \times k$, $\alpha \in [0,1]$.  For plate like objects ($1/2 < \alpha <1$), it is possible to show rigorously, for $k \gg 1$, the existence of a uniaxial nematic phase, where only minor axes of plates are aligned parallel to each other, and  there is no translational order~\cite{2018-dgj-araiv-plate}. However, the behavior of the corresponding lattice model, which is also interesting in connection with certain resonating plaquette wavefunctions and the possibility of a lattice realization of a liquid state of fluctuating quadrupoles~\cite{pankov2007resonating}, has not been studied away from full-packing.

With this motivation, here we study the phase diagram of a system of $2\times 2\times 1$ hard plates on the three dimensional cubic lattice, {\em i.e.} a lattice gas of plates that each cover an elementary plaquette of the cubic lattice and occupy its four vertices, with the constraint that no two plates occupy the same site of the cubic lattice. We use a cluster algorithm and focus here on the isotropic system, with equal fugacities for the three orientations of plates, so that ``$\mu$-type plates'' (with normal along the $\mu$ axis) have equal fugacity for all $\mu$ ($\mu = x, y, z$) (see Ref.~\cite{Geetpaper} for the anisotropic fully-packed case in which every site of the cubic lattice is occupied by exactly one plate). We show, using grand canonical Monte Carlo simulations, that the system undergoes two phase transitions as a function of increasing fugacity: first from a disordered fluid to a spontaneously layered phase,  and second from this layered phase to a sublattice ordered phase. 

In the layered phase, the system breaks up into disjoint slabs of thickness two along one spontaneously chosen cartesian direction. Plates with normals perpendicular to this layering direction are preferentially contained entirely within these slabs,  while plates straddling two successive slabs have a lower density. This corresponds to a two-fold symmetry breaking of translation symmetry along one spontaneously chosen cartesian direction, leading to ``occupied slabs'' stacked along the layering direction with a separation of one lattice spacing.  Additionally the symmetry between the three types of plates is spontaneously broken, as plates with normal along the layering direction have a lower density than the other two types of plates. Intriguingly, the occupied slabs exhibit two-dimensional power-law columnar order. In contrast, inter-slab correlations of the two-dimensional columnar order parameter decay exponentially with the separation between the slabs. In addition, the layered phase breaks  the symmetry between the three types of plates: plates with normal along the layering direction have a lower density than the other two types of plates. 

\begin{figure}
\includegraphics[width=\columnwidth]{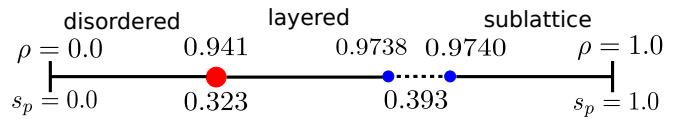}
\caption{Schematic phase diagram of $2\times 2\times 1$ hard plates model. The red dot represents a continuous transition and the blue dots and dotted line represent the coexistence regime in  a first order transition.
\label{schematic}}
\end{figure}

In the sublattice ordered phase, there is two-fold ($Z_2$) breaking of lattice translation symmetry along all three cartesian directions.  In this phase,  the corner of a $\mu$-type plate with the smallest $\nu$ coordinates (for both $\nu \neq \mu$) preferentially occupies one spontaneously chosen sublattice out of the eight sublattices of vertices of the cubic lattice, and each type of plate breaks translational symmetry along the two directions perpendicular to its normal (see Sec.~\ref{sec:phases} for a more detailed description). The disordered to layered transition occurs at density $\rho^{DL}\approx 0.941$.  From finite size scaling, we show that this transition is continuous, with properties that are consistent with those of the $O(3)$ universality class perturbed by cubic anisotropy.   The transition from layered to sublattice phase occurs at density $\rho^{LS}\approx 0.974$. We show that this second transition is first-order. The overall structure of the phase diagram found here is summarized in Fig.~\ref{schematic}.

Finally, we note that the fully packed system of $2\times 2\times 1$ hard plates on the cubic lattice also has a very rich phase diagram as a function of anisotropy in the fugacity of the three orientation of plates. This is discussed in a parallel work~\cite{Geetpaper}.


\section{\label{sec:model}Model and algorithm}

Consider a $L \times L \times L$ cubic lattice with periodic boundary along the three orthogonal directions. The lattice sites may be empty or occupied by $2 \times 2 \times 1$  plates, each of which covers an elementary plaquette of the cubic lattice and occupies the four sites of the corresponding plaquette. Three types of plates are possible depending on the orientation of the normal to the plate, i.e., $x$, $y$ and $z$ plates corresponding to plates lying in the $yz$, $zx$ and $xy$ planes respectively. The plates interact through a hard-core constraint, i.e., no two plates
may occupy the same site of the cubic lattice. We associate activity $s_p$ and $s_0$ to each  plate and vacancy respectively. These are normalized through
\be
s_p^{1/4}+s_0=1,
\ee
where the power $1/4$ accounts for the fact that a plate touches four vertices, while a vacancy resides on one vertex.

We study the system using grand canonical Monte Carlo simulations. Conventional Monte Carlo simulations involving local evaporation, deposition, diffusion, and rotation moves are inefficient in equilibrating such systems especially when the packing fraction approaches full packing. These difficulties may be over come by algorithms that include cluster moves. The transfer-matrix algorithm we use updates strips of sites of size proportional to $L$. This has been particularly useful in earlier studies of other hard core lattice gas models~\cite{2012-krds-aipcp-monte,2013-krds-pre-nematic,2015-rdd-prl-columnar,2014-nr-pre-multiple}. 
Below we provide a brief description of this algorithm, and give details of its implementation for our system of hard plates; we will follow terminology of Ref.~\cite{2015-rdd-prl-columnar}, where the phase diagram was obtained for a mixture of dimers and squares on a square lattice at all packing densities using such a transfer-matrix algorithm. 

\begin{figure}
\includegraphics[width=\columnwidth]{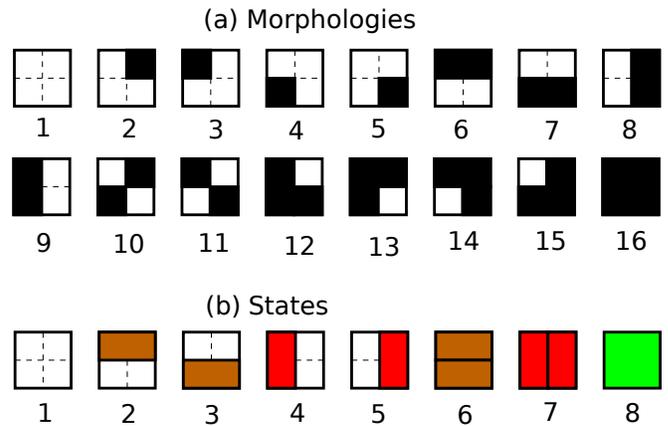}
\caption{Schematic diagram of (a) sixteen possible morphologies and (b) eight possible states of the $2 \times 2 \times L$ tube, used to construct the transfer matrix. To represent different states we have taken the projection in $xy$-plane. Black represent blocked site and brown, red, green respectively represents projection of $y$, $x$ and $z$ plates. Note that a vertex of the original cubic lattice is represented by an elementary cube in this space-filling representation for ease of visualization, and the morphologies and states are then depicted in terms of a cross-sectional view of the tube.}
\label{states_label}
\end{figure}

We define a ``tube'' to be a cuboidal subset of the $L \times L \times L$ lattice, of size $2\times2\times L$  and made up of $L$ plaquettes of size $2\times 2\times 1$ stacked along one cartesian axis.
Choose a tube at random in any one of the three orthogonal directions. 
Remove all the plates that are completely contained within the tube. There may be some protruding plates that are not fully contained within the tube, but touch sites of this tube.  These plates are left undisturbed.  Due to these protrusions, the shape of the tube (after removal of fully contained plates) is complicated and can
be characterized by assigning different morphologies to each section depending on the protrusion. There are are $16$ such morphologies possible for 
each section and they are listed in Fig.~\ref{states_label}(a). In order to provide a visual depiction that is easier to read, we use  a space-filling convention for depicting the protruding plates. In this space filling convention, each site of the original cubic lattice maps to a unit cube of the dual cubic lattice, and each plate is a space-filling object that occupies a $2\times 2\times 1$ slab consisting of 4 adjacent elementary cubes of the dual lattice. Note that this alternate description is behind the commonly used terminology, also used here, which refers to the hard plates as $2\times 2\times 1$ cuboids.

The aim is to refill the tube with a new configuration of plates that are fully contained within the tube, but with the correct equilibrium probability. The probability of this new configuration may be calculated using transfer matrices. 
Any $2\times2\times 1$ section with a given morphology may be filled by plates in at most eight different ways. The possible states for a section are listed in Fig.~\ref{states_label}(b). Among the sixteen possible morphologies, there are fifteen morphologies with partially blocked sites. The remaining one morphology [morphology-$16$ as shown in Fig.~\ref{states_label} (a)] represents a complete blockage in the chosen tube. We have to thus calculate $15^2=225$ different transfer matrices of size $8\times 8$. Let $T_{m_1,m_2}$ be the transfer matrix where the system is transferring from morphology $m_2$ to morphology $m_1$. The matrix element may be written as
\be
T_{m_1,m_2}(i,j)=c_{m_1,m_2}(i,j)W_pW_0,
\ee
where $c_{m_1,m_2}(i,j)$ is the compatibility factor, $W_p$ is the weight associated with the particle that sits on morphology $m_1$ and $W_0$ is the weight of vacancies present  on morphology $m_2$ after depositing particle on morphology $m_1$. The compatibility factor $c_{m_1,m_2}(i,j)$ is $1$ if the states $i$ and $j$ are compatible on morphologies $m_1$ and $m_2$, otherwise it equals zero. The weights associated with the particles and vacancies may be written as
\bea
W_p&=&s_p^{n_s}, ~n_s=0,1,2,\\
W_0&=&s_0^{n_0}, ~n_0=0,1,2,3,4.
\eea

Examples of few transfer matrices are given in Eqs.~(\ref{t11})--(\ref{t31}).
\be
\label{t11}
T_{1,1} = 
\left(\begin{array}{cccccccc} 
s_0^4   &   s_0^2 & s_0^2 & s_0^2 & s_0^2 & 1 & 1 & 1 \\
s_ps_0^2 & 0 & s_p &0 &0 &0 &0 &0 \\
s_ps_0^2 & s_p & 0 &0 &0 &0 &0 &0 \\
s_ps_0^2 & 0 & 0 &0 &s_p &0 &0 &0 \\
s_ps_0^2 & 0 & 0 &s_p &0 &0 &0 &0 \\
s_p^2 & 0 & 0 &0 &0 &0 &0 &0 \\
s_p^2 & 0 & 0 &0 &0 &0 &0 &0 \\
s_ps_0^4   &   s_ps_0^2 & s_ps_0^2 & s_ps_0^2 & s_ps_0^2 & s_p & s_p & s_p 
\end{array}\right)
\ee
\be
\label{t13}
T_{1,3} = 
\left(\begin{array}{cccccccc} 
s_0^3   &   0 & s_0 & 0 & s_0 & 0 & 0 & 0 \\
0 & 0 & 0 &0 &0 &0 &0 &0 \\
s_ps_0 & 0 & 0 &0 &0 &0 &0 &0 \\
0 & 0 & 0 &0 &0 &0 &0 &0 \\
s_ps_0 & 0 & 0 &0 &0 &0 &0 &0 \\
0 & 0 & 0 &0 &0 &0 &0 &0 \\
0 & 0 & 0 &0 &0 &0 &0 &0 \\
s_ps_0^3   &   0 & s_ps_0 & 0 & s_ps_0 & 0 & 0 & 0 
\end{array}\right)
\ee
\be
\label{t31}
T_{3,1} = 
\left(\begin{array}{cccccccc} 
s_0^4   &   s_0^2 & s_0^2 & s_0^2 & s_0^2 & 1 & 1 & 1 \\
0 & 0 & 0 &0 &0 &0 &0 &0 \\
s_ps_0^2 & s_p & 0 &0 &0 &0 &0 &0 \\
0 & 0 & 0 &0 &0 &0 &0 &0 \\
s_ps_0^2 & 0 & 0 &s_p &0 &0 &0 &0 \\
0 & 0 & 0 &0 &0 &0 &0 &0 \\
0 & 0 & 0 &0 &0 &0 &0 &0 \\
0 & 0 & 0 &0 &0 &0 &0 &0 
\end{array}\right)
\ee

The partition function of a closed $2\times2\times L$ tube with morphology $m_1,\dots,m_L$ may be written as
\be
Z^c=\sum_{i}\langle i|T_{m_L,m_1}T_{m_1,m_2}\dots T_{m_{L-1},m_L}|i\rangle,
\ee
where $|i\rangle$ is the state vector of state $i$.
The partition function for the open tube of length $X<L$ may be written as
\be
Z^o=\langle \mathcal{L}_{m_1}|T_{m_1,m_2}T_{m_2,m_3}\dots T_{m_{X-1},m_X}|\mathcal{R}_{m_X}\rangle,
\ee
where $\langle \mathcal{L}_{m_1}|$ and $|\mathcal{R}_{m_X}\rangle$ are respectively left and right vectors that may be written as
\bea
\mathcal{L}_{m_1}(n) &=&T_{16,m_1}(1,n),\\
\mathcal{R}_{m_X}(n) &=&T_{m_X,16}(n,1).
\eea
Calculating the partition function, we re-occupy the tube, section by section, according to the calculated probabilities. Disjoint tubes are updated simultaneously in our parallelized implementation.
To speed up equilibration as well as to reduce autocorrelation times, we also implement a flip move in which a pair of adjacent parallel plates of same type is replaced by another pair of adjacent parallel plates whose type is chosen randomly.  For each value of activity, we ensure that equilibration has been achieved by starting the simulations with configurations that correspond to different phases, and ensuring that the final equilibrium state is independent of the initial state. 

\section{\label{sec:phases}Different Phases of system}

\subsection{\label{subsec:characterisation} Observables and order parameters}

\begin{figure}
\includegraphics[width=\columnwidth]{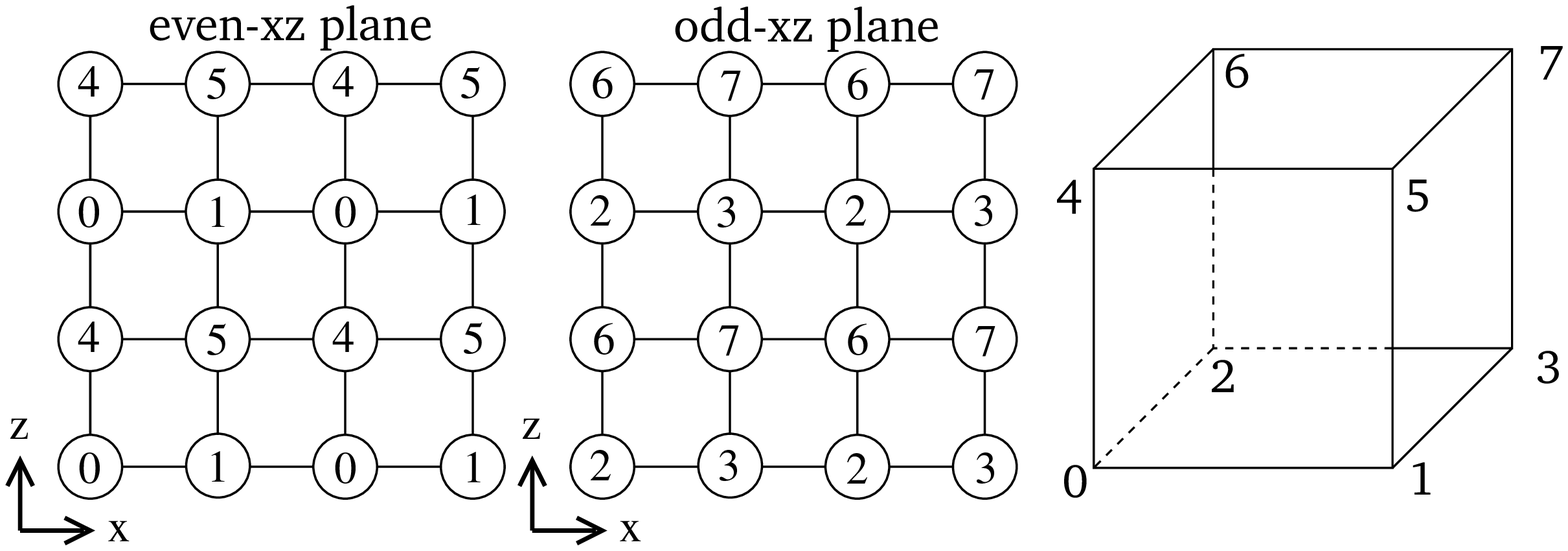}
\caption{Division of the full lattice into eight sublattices $0, 1, \dots, 7$, depending on whether each coordinate is odd or even. The arrows show the orientation of the  three axes $x$, $y$ and $z$.}
\label{sublat_label}
\end{figure}

As the density is varied, we observe three different phases in our simulations. To characterize them, it is convenient to 
divide the full lattice into eight sublattices depending on whether the $x$, $y$, and $z$ coordinates of a  site are even (0) or odd (1), 
as shown in Fig.~\ref{sublat_label}. A lattice site $(x,y,z)$ belongs to the sublattice constructed out of the binary number $zyx$ where each of the digits is the corresponding coordinate taken modulo two. Except for plates that cover a plaquette on an edge that wraps around the periodic direction, we assign each plate to the site with least $x$, $y$ and $z$ coordinates (of the four sites touched by it). For plates on wrapping plaquettes, this definition is of course modified in the obvious way to remain consistent with the treatment of bulk plates. The corner that occupies this site to which a plate ``belongs'' is the ``head'' of the plate.

To characterize the phases quantitatively, we define sublattice densities $\rho_i^j$ as the volume fraction of plates of type $j=x,y,z$ whose heads occupy sites of sublattice  $i=0,\ldots,7$. We also define three particle densities $\rho^j$, eight sublattice densities $\rho_i$, and total density $\rho$ as
\begin{align}
\rho^j & =\sum_{i=0}^7 \rho_i^j,~j=x,y,z, \nonumber \\
\rho_i & =\sum_{j=x,y,z}\rho_i^j,~i=0,\dots,7, \\
\rho &= \sum_{i=0}^7 \rho_i. \nonumber
\end{align}

To quantify the breaking of translational invariance in the different directions,  it is convenient to define the quantities
\bea
\ell_x&=&\frac{1}{L^3}\sum_{x,y,z}\phi(x,y,z)(-1)^x,\nonumber\\
\ell_y&=&\frac{1}{L^3}\sum_{x,y,z}\phi(x,y,z)(-1)^y,\\
\ell_z&=&\frac{1}{L^3}\sum_{x,y,z}\phi(x,y,z)(-1)^z,\nonumber
\eea
where $\phi(x,y,z)$ is $1$ if the site is occupied by the head of a plate and zero otherwise. 
The square of the layering order parameter, which characterizes the layered phase, may be defined as
\be
\Lambda^2=\ell_x^2+\ell_y^2+\ell_z^2.
\label{eq:lambda_sq}
\ee
The columnar vector $\vec{C}$ with components $({c}_x, {c}_y, {c}_z)$ may be written as
\bea
{c}_x&=&\frac{1}{L^3}\sum_{x,y,z}(-1)^{y+z}\phi(x,y,z),\nonumber\\
{c}_y&=&\frac{1}{L^3}\sum_{x,y,z}(-1)^{x+z}\phi(x,y,z),\\
{c}_z&=&\frac{1}{L^3}\sum_{x,y,z}(-1)^{x+y}\phi(x,y,z).\nonumber
\eea
The square of the columnar order parameter may be defined as
\be
\Gamma^2={c}_x^2+{c}_y^2+{c}_z^2.
\ee
We also define the square of the order parameter $\omega$ to characterize the sublattice phase
\be
\omega^2=\ell_x^2\ell_y^2\ell_z^2.
\ee
To capture the breaking of particle number symmetry, we define a nematic order parameter $\Pi$ as
\be
\Pi^2=\big({\rho^z-\frac{\rho^y}{2}-\frac{\rho^x}{2}}{\big)}^2+\frac{3}{4}\big({\rho^y}-{\rho^x}{\big)}^2.
\label{eq:qn}
\ee
When $\Pi$ is non-zero,  particle symmetry is broken.

In a sublattice ordered phase, we expect $\omega^2$, $\Lambda^2$ and $\Gamma^2$ to all tend to nonzero values in the thermodynamic limit. In contrast, in the layered phase, we expect $\omega^2$ to tend to zero as $1/L^6$ and $\Gamma^2$ to tend to zero as $1/L^3$ in the thermodynamic limit, while $\Lambda^2$ tends to a nonzero limit.

\begin{figure*}
\includegraphics[width=2.0\columnwidth]{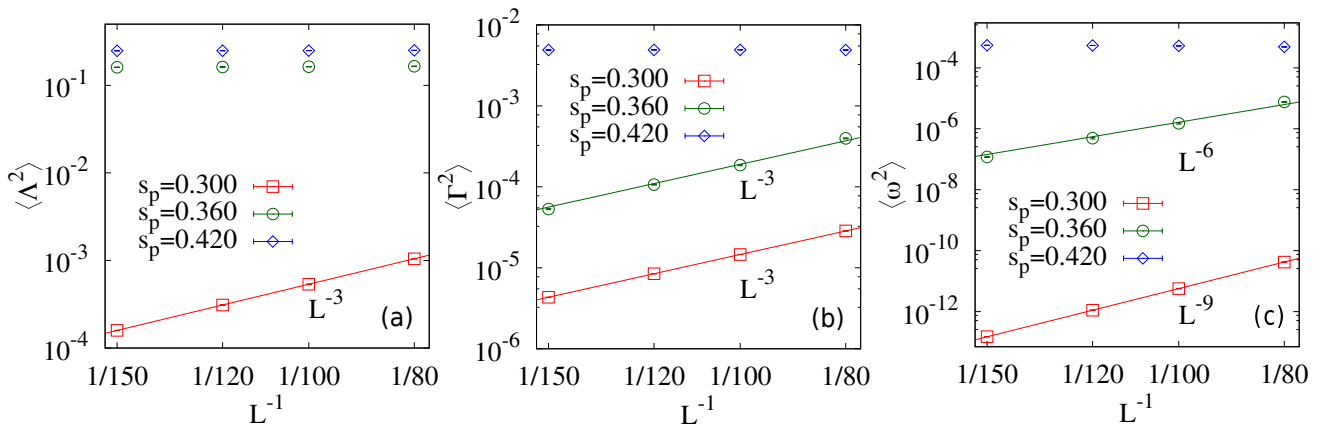}
\caption{Variation of the square of the (a) layered order parameter $\langle\Lambda^2 \rangle$, (b) columnar order parameter $\langle\Gamma^2 \rangle$ and (c) sublattice order parameter $\langle\omega^2 \rangle$ as a function of $L^{-1}$ for different values of $s_p$.}
\label{fig:op_sq_Linv}
\end{figure*}

In Fig.~\ref{fig:op_sq_Linv}(a-c), we display $\Lambda^2$, $\Gamma^2$ and $\omega^2$ as a function of $L^{-1}$ for $s_p=0.300$, $s_p=0.360$ and $s_p=0.420$. The quantity $\Lambda^2$ decays to zero as $L^{-3}$ for $s_p=0.300$ and takes non-zero values for $s_p=0.360$ and $s_p=0.420$. The quantity $\Gamma^2$ decays to zero as $L^{-3}$ for both $s_p=0.300$ and $s_p=0.360$, and takes non-zero value for $s_p=0.420$. Similarly $\omega^2$ also decays to zero for both $s_p=0.300$ and $s_p=0.360$, but the decay obey different power laws, which are $L^{-9}$ and $L^{-6}$ respectively. For $s_p=0.420$, $\omega^2$ takes non-zero values. Taken together, these behaviours allow us to conclude that the system is successively in a disordered, layered and sublattice-ordered phase for $s_p=0.300$, $s_p=0.360$, and $s_p=0.420$ respectively. This establishes the presence of the three phases described in our introductory discussion. 

\begin{figure}
\includegraphics[width=\columnwidth]{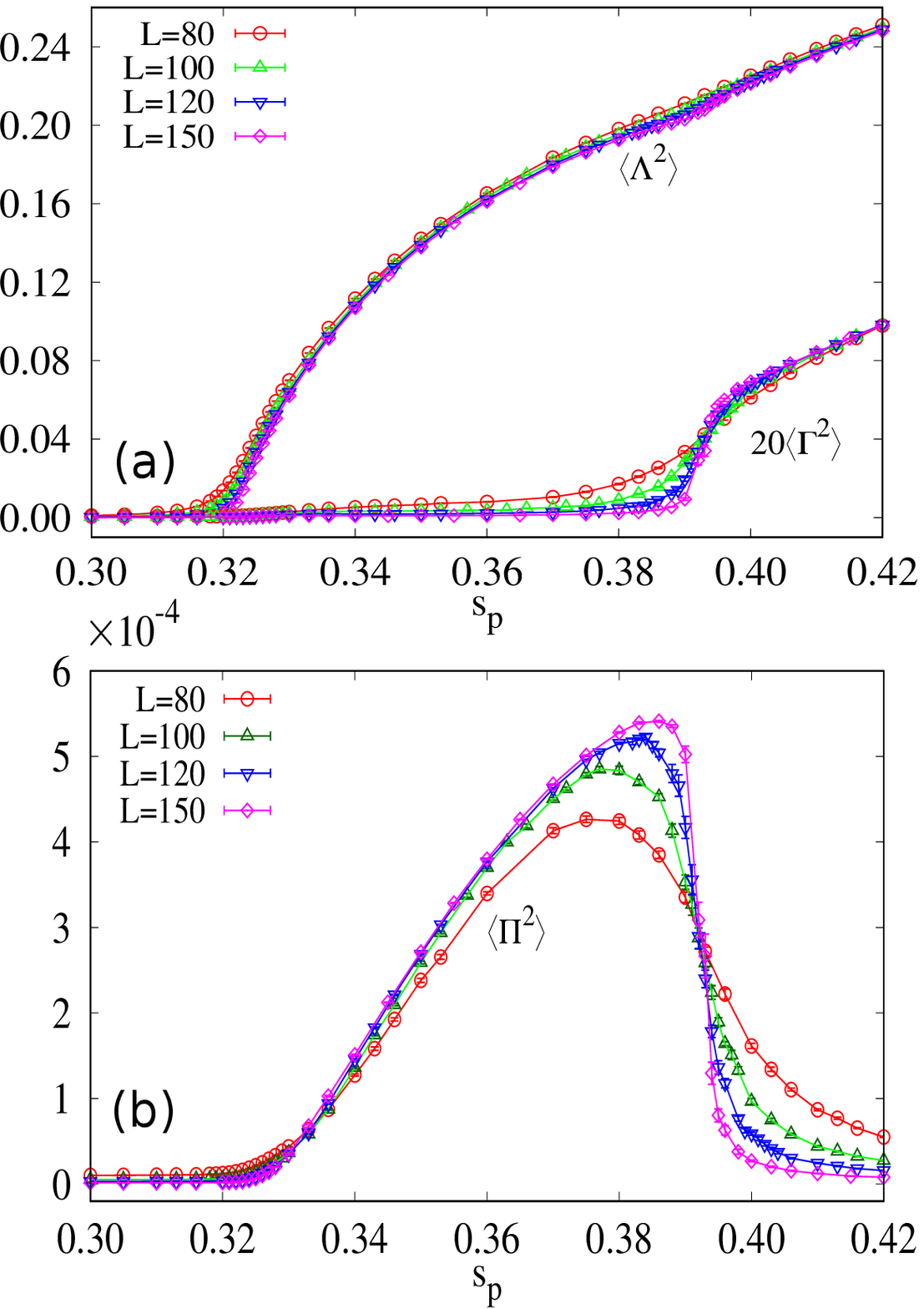}
\caption{Variation of the square of the (a) translational order parameters $\langle\Lambda^2\rangle$, $\langle\Gamma^2\rangle$, and (b) nematic order parameter $\langle\Pi^2\rangle$  with activity of plate $s_p$. The data are for  for system sizes $L=80, 100, 120, 150$.}
\label{combined_translational}
\end{figure}

For a bird's eye view of the phase diagram as a function of plate fugacity, we plot the fugacity dependence of the various order parameters in Fig.~\ref{combined_translational}.  We clearly observe a layered phase ($\Lambda^2 \neq 0$, $\Gamma^2=0$, $\omega^2=0$) and a sublattice phase ($\Lambda^2 \neq 0$, $\Gamma^2\neq 0$, $\omega^2\neq0$). The variation of $\Pi^2$ as a function of $s_p$ is also shown in Fig.~\ref{combined_translational}. $\Pi^2$ is zero in both disordered and sublattice phase, and  takes nonzero values only in the layered phase, which indicates asymmetric densities of three types of particles in the layered phase.

\subsection{Disordered phase}
The characterization of the disordered phase is straightforward. All order parameters vanish in the thermodynamic limit in this  low density phase. The plates form a disordered fluid, with their heads uniformly distributed, i.e. each of the sublattice densities are equal for the three different types of plates, {\em i.e.},
\bea
\rho_i &=& \frac{\rho}{8}, \quad i=0, 1, \ldots, 7,\nonumber \\
\rho^j &=& \frac{\rho}{3}, \quad j=x, y, z \; .  \nonumber
\eea

\subsection{\label{subsec:layered} Layered phase}

\begin{figure}
\includegraphics[width=\columnwidth]{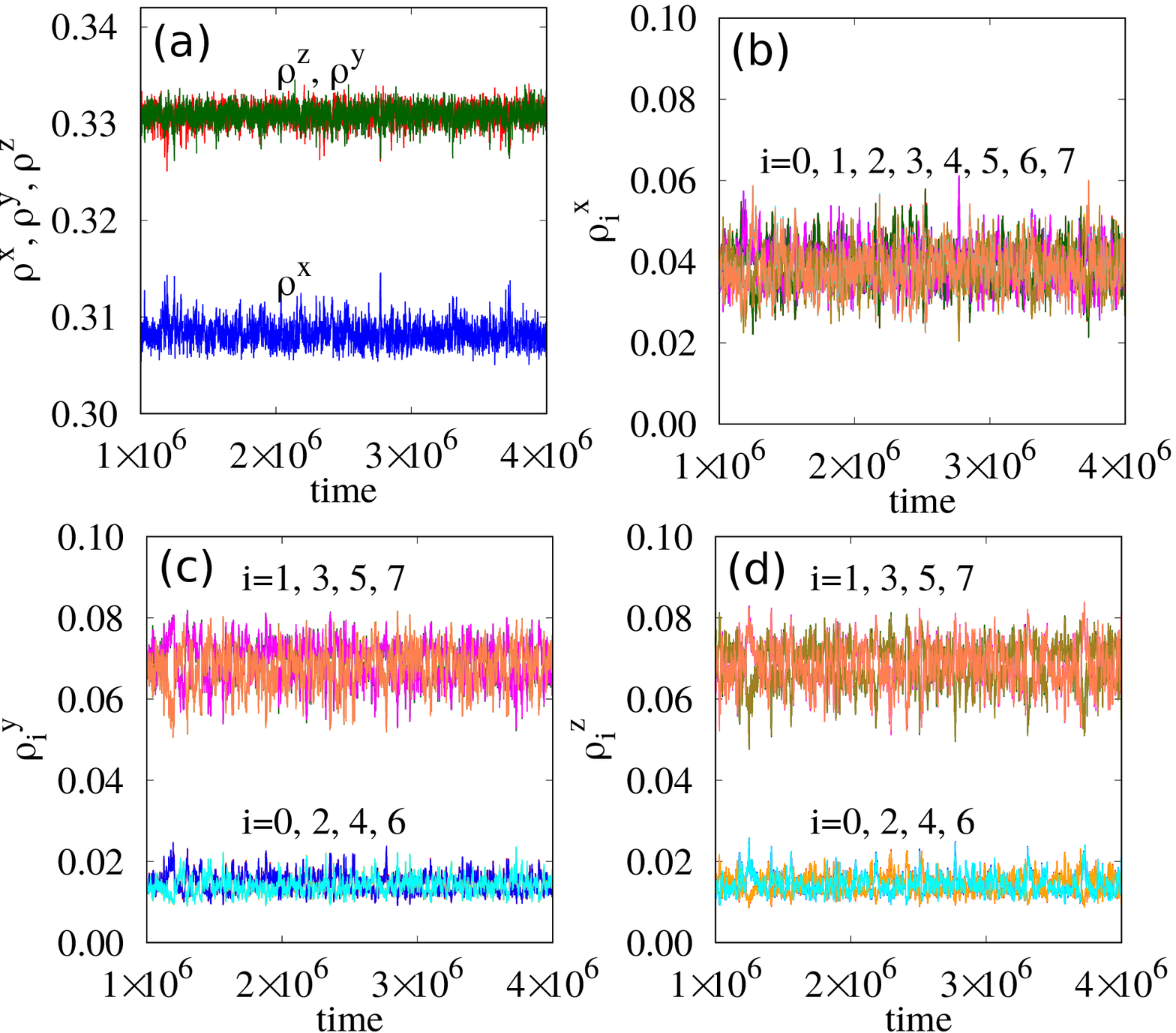}
\caption{The temporal evolution of different thermodynamic quantities is shown for an equilibrated layered phase at  activity  $s_p=0.380$ and for system size $L=120$. (a) The three plate densities $\rho^x, \rho^y, \rho^z$. The eight sublattice densities $\rho_i$  for  (b) $x$-plates, (c) $y$-plates,  and (d) $z$-plates, where the subscripts $i=0,\dots,7$ denote the different sublattices and the superscripts $x$, $y$, $z$ denote the different types of plates.}
\label{time_pf_lay}
\end{figure}

With increasing density, we observe that the system undergoes a transition into the  layered phase described in the Introduction. 
In Fig.~\ref{time_pf_lay}, we display the time evolution of the sublattice densities when the system is in a layered phase with layering in $x$-direction. Fig.~\ref{time_pf_lay}(a) compares the densities of the three types of plates. It is clear that the density of $x$-plates is suppressed compared to $y$ and $z$-plates, when the layering is in the $x$-direction, i.e., $\rho^y\approx \rho^z \gg \rho^x$. At the same time, Fig.~\ref{time_pf_lay}(b)--(d) show that while the heads of $x$-plates occupy all sublattices equally, the heads of $y$ and $z$-plates preferentially occupy planes with odd $x$ (in this case), contributing to $\rho_1$, $\rho_3$, $\rho_5$, and $\rho_7$. These observations lead us to the basic picture of the layered phase described earlier in the Introduction. Evidence for power-law columnar order within the occupied slabs is discussed separately in Sec.~\ref{sec:layered}.

\subsection{\label{subsec:sublattice} Sublattice-ordered phase}

At higher densities including full packing, we observe a sublattice-ordered phase.  In this phase, all three types of plates are equivalent, but translational invariance is broken in all three directions, as in a solid. 

\begin{figure}
\includegraphics[width=\columnwidth]{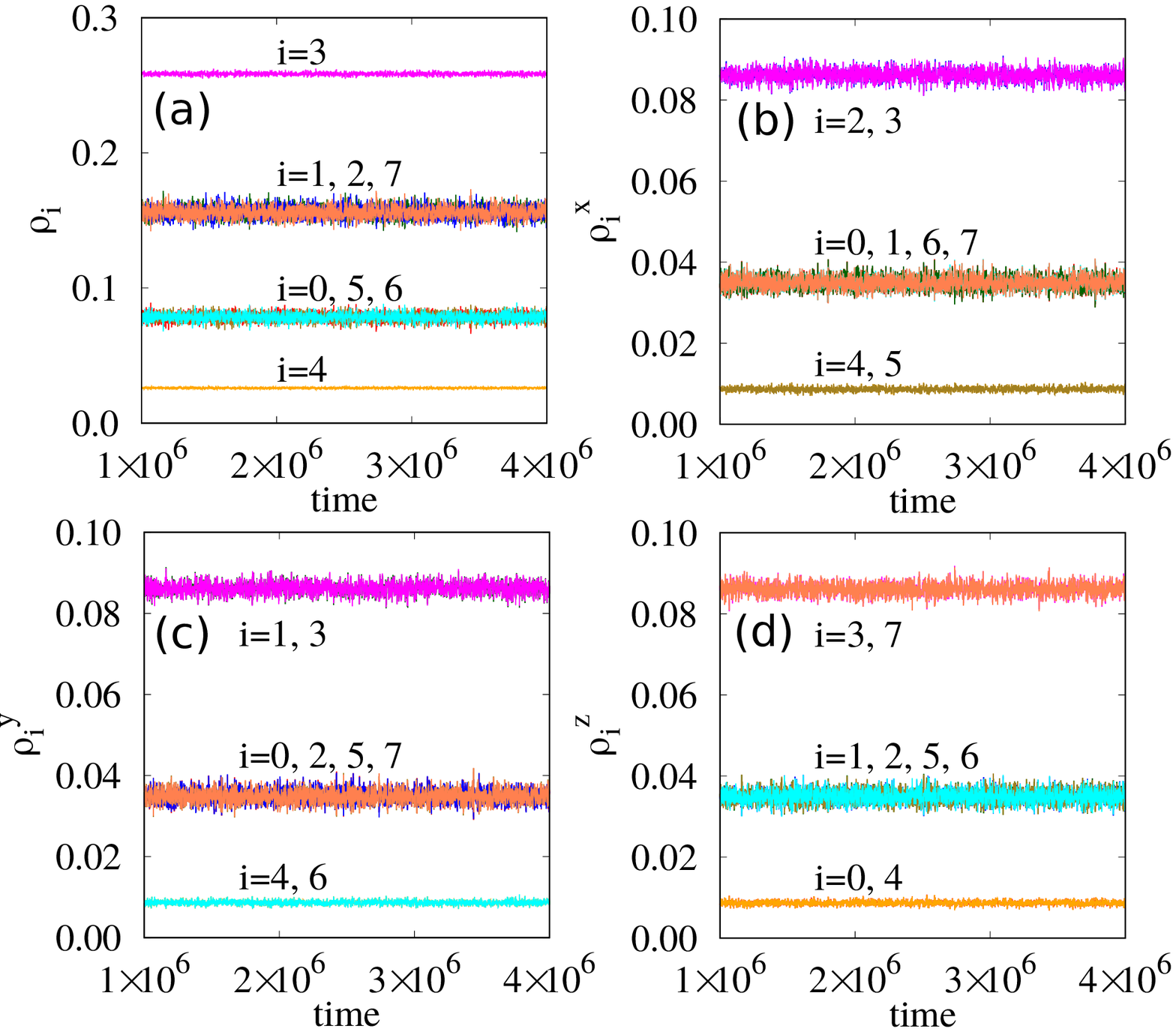}
\caption{The temporal evolution of different thermodynamic quantities is shown for an equilibrated sublattice phase at  activity  $s_p=0.460$ and for system size $L=120$. The eight sublattice densities (a) $\rho_i$ summing over three types of particles, for individual particle type (b) $x$-plates, (c) $y$-plates and (d) $z$-plates, where the subscripts $i=0,\dots,7$ denote the different sublattices and the superscripts $x$, $y$, $z$ denote the different types of plates.}
\label{time_pf_sub}
\end{figure}
In Fig.~\ref{time_pf_sub}, we display the time evolution of the sublattice densities when the system is in a sublattice-ordered phase. Out of the eight sublattices, one of them is occupied preferentially. At the same time, there is a solid-like sublattice ordering as can be seen seen from Fig.~\ref{time_pf_sub}(a). The sublattice densities for each type of plate are shown in Fig.~\ref{time_pf_sub}(b)--(d). For each type of the plates, two sublattices are preferred, as in a columnar phase. The preferred sublattice densities are $[\rho_2, \rho_3]$, $[\rho_1, \rho_3]$ and $[\rho_3, \rho_7]$ for $x$, $y$ and $z$-plates respectively. Time profile of total sublattice density $\rho_i$ breaks into four labels [see Fig.~\ref{time_pf_sub}(a)]. The top and bottom labels are $\rho_3$ and $\rho_4$ respectively. Two intermediate labels are degenerate with three densities in each label. Higher intermediate label has densities $\rho_1, \rho_2, \rho_7$ and lower intermediate label has densities $\rho_0, \rho_5, \rho_6$ respectively. The pattern of the labels may be understood from the right panel of Fig.~\ref{sublat_label} where the sublattice division is shown schematically. The labels are divided depending on the lowest distance between the sublattice-$3$ (most occupied) and other sublattices. The density decreases with increasing distance between sublattices.

One could imagine the sublattice phase as follows. Consider a collection of $2\times 2 \times 2$ cubes that are arranged in a periodic manner to favor one sublattice. If the cubes are now replaced by a pair of plates of the same kind (each cube can thus be replaced by parallel plates in three ways), then the phase that is obtained is similar to the sublattice phase that we see in the system of hard plates. Unlike the layered phase, the densities of the three types of plates are equal. 
For the fully packed case, this picture gives a lower bound to the entropy per site of $(1/8)\log(3)$.

\section{\label{sec:transitions}Phase transitions}
We now study the nature of the two phase transitions we observe, from the disordered to layered phase, and from the layered to sublattice-ordered phase. 

\subsubsection{Disordered to layered phase transition}

As noted already, it is convenient to focus on the squared order parameter $\Lambda^2$ as defined in Eq.~(\ref{eq:lambda_sq}) to probe the symmetry breaking accompanying the layering transition. In the disordered phase, $\Lambda^2 \to 0$ in the thermodynamic limit, while the thermodynamic limit of $\Lambda^2$ in the layered phase is nonzero. As we have already seen in Fig.~\ref{combined_translational}(a), the first transition encountered as one increases the activity $s_p$ from small values is a transition from a disordered to a layered phase signalled by a threshold at which $\Lambda^2$ develops a nonzero value in the thermodynamic limit.

\begin{figure}
\includegraphics[width=\columnwidth]{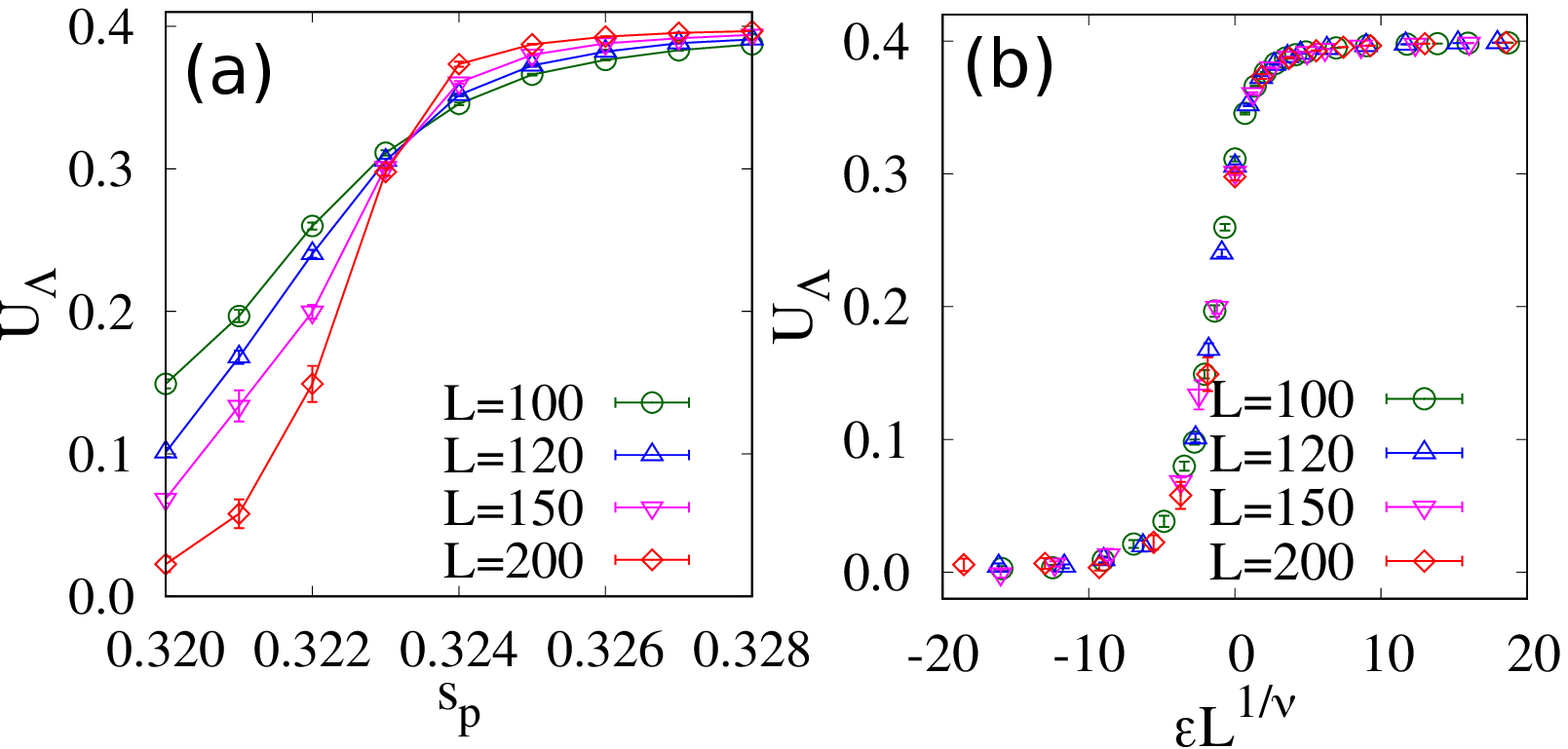}
\caption{Data for Binder cumulant $U_{\Lambda}$ near the disordered-layered transition. (a) $U_{\Lambda}$ for different system sizes intersect close to $s_p^{DL} \approx 0.323$. (b) $U_{\Lambda}$ for different system sizes collapse onto a curve when the parameter are scaled as in Eq.~(\ref{eq:uscaling}) with exponent $\nu=0.704$.}
\label{transition_1}
\end{figure}

For a more detailed understanding of the disordered-layered transition, we also measure the Binder cumulant $U_{\Lambda}$ associated with $\Lambda^2$
\be
U_\Lambda=1-\frac{9}{15}\frac{\langle \Lambda^4\rangle}{\langle \Lambda^2\rangle ^2}.\label{eq:define_binder}
\ee
From standard finite-size scaling theory of continuous phase transitions, we expect that
the Binder cumulant obeys a scaling form near the critical point:
\be
U_\Lambda(\epsilon, L) \simeq f_\Lambda (\epsilon L^{1/\nu}),\label{eq:uscaling}
\ee
where $\epsilon=s_p-s_c$ is the deviation from the critical point, $\nu$ is the critical exponent, and $f_\Lambda$ is the scaling function. The nature of the symmetry breaking associated with the layering transition suggests that the finite-size scaling of this Binder cumulant should be governed by the scaling behavior in the O(3) universality class with cubic anisotropy, wherein  the values of the critical exponents are known to be $\nu=0.704$, $\beta=0.362$, and $\gamma=1.389$ ~\cite{carmona-2000,caselle-1998}. 

The variation of $U_\Lambda$ with $s_p$ for different system sizes is shown in Fig.~\ref{transition_1}(a). The data for different system sizes cross each other at the critical point  $s_p^{DL} \approx 0.323$. The corresponding critical density is $\rho^{DL}\approx0.940$.
These data for the Binder cumulant for different $L$ collapse to a reasonable accuracy onto a single scaling curve when the variables $\epsilon$ is scaled as in
Eq.~(\ref{eq:uscaling}) with the theoretical value of $\nu=0.704$, as shown in Fig.~\ref{transition_1}(b). The quality of this data collapse bears out our initial theoretical expectation that the transition is in the universality class of the  three dimensional Heisenberg model with cubic anisotropy~\cite{carmona-2000,caselle-1998}. 

\subsubsection{Layered to sublattice phase  transition}

In this section, we study the nature of the second transition from layered to sublattice phase. Suitable order parameters are  $\Gamma^2$ and $\Pi^2$ as defined in Eq.~(\ref{eq:lambda_sq}) and Eq.~(\ref{eq:qn}) respectively. The associated Binder cumulants  may be defined as
\bea
U_\Gamma&=&1-\frac{1}{2}\frac{\langle \Gamma^4\rangle}{\langle \Gamma^2\rangle ^2},\label{eq:binder_gamma}\\
U_\Pi&=&1-\frac{1}{2}\frac{\langle \Pi^4\rangle}{\langle \Pi^2\rangle ^2},\label{eq:binder_pi}
\eea 
We show that the transition is first-order in nature.

\begin{figure*}
\includegraphics[width=2.0\columnwidth]{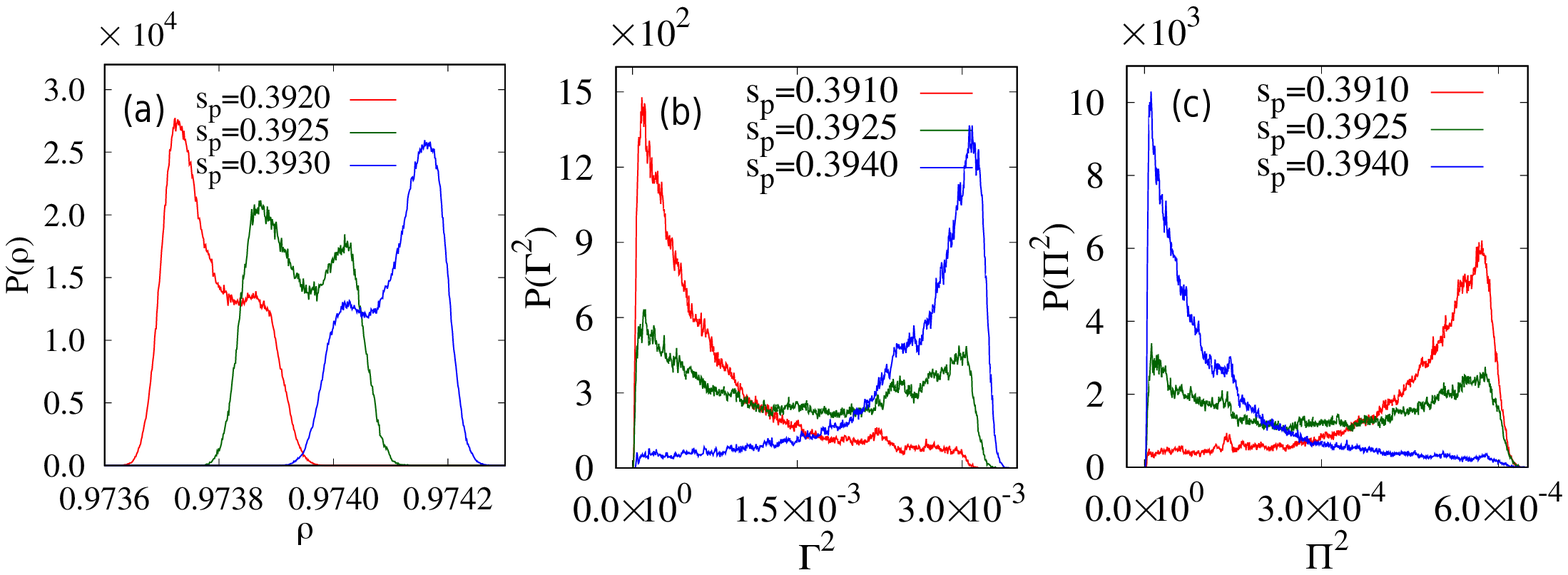}
\caption{Plot of probability distribution of (a) total density $\rho$, (b) $\Gamma^2$ and (c) $\Pi^2$ near layered to sublattice transition for $L=150$.}
\label{transition_2_hist}
\end{figure*}

\begin{figure}
\includegraphics[width=\columnwidth]{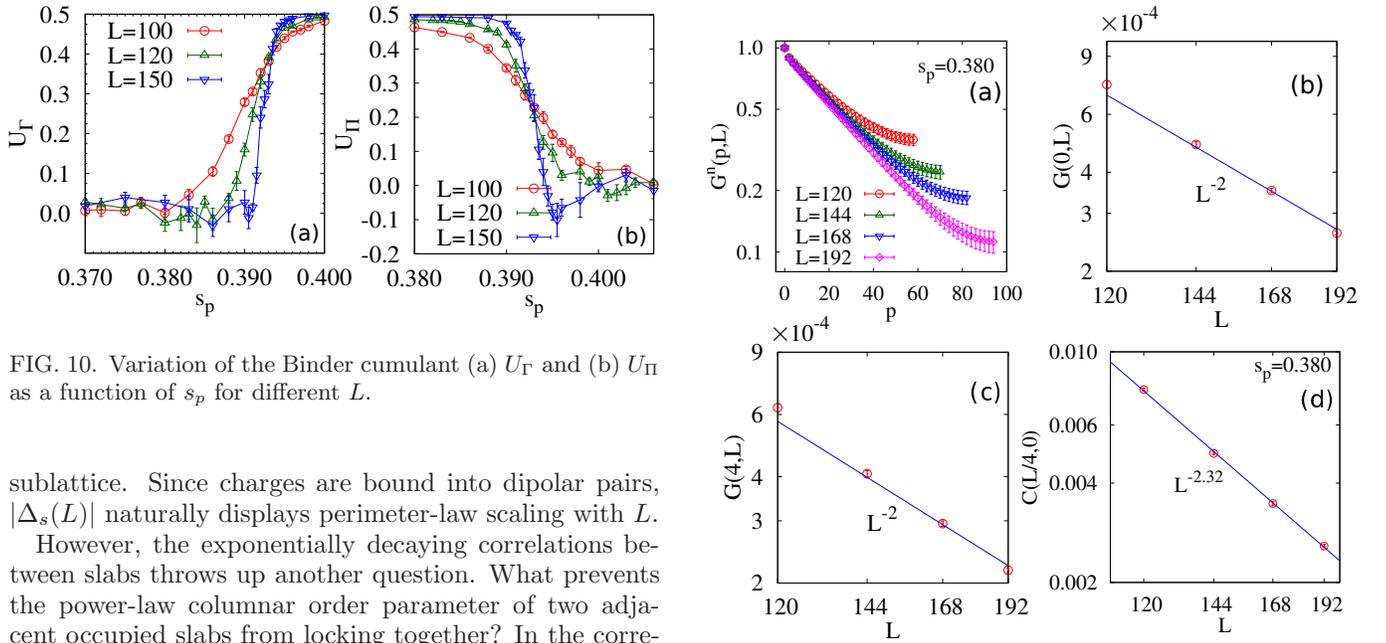}
\caption{Variation of the Binder cumulant (a) $U_{\Gamma}$ and (b) $U_{\Pi}$ as a function of $s_p$ for different $L$.}
\label{transition_2_binder}
\end{figure}

The variation of $\langle \Gamma^2\rangle$ and $\langle \Pi^2\rangle$ with $s_p$, for different system sizes, has already been displayed in Fig.~\ref{combined_translational}(a) and (b) respectively; both order parameters have a sharp variation across the transition point, and the data for different system sizes intersect each other with the curves becoming steeper with increasing system size. These are signatures of a first-order transition. In Fig.~\ref{transition_2_hist}(a), (b) and (c), we also display the measured histograms of the total plate density $\rho$ and order parameters $\Gamma^2$, $\Pi^2$ . To increase our signal to noise ratio, we easure these histograms by averaging the time series of each observable over a bin of $51$ successive measurements and then recording the histogram of the resulting bin averages. As is clear from this figure, these histograms have a double-peaked distribution characteristic of phase coexistence at a first-order transition. However, the jump in the density across the transition is quite small (of the order $10^{-4}$) and therefore quite difficult to detect directly in simulations. More evidence in support of the first-order nature of this transition is provided by the Binder cumulants of the order parameters. The variation of Binder cumulants $U_\Gamma$ and $U_\Pi$ is shown in Fig.~\ref{transition_2_binder}(a) and (b) respectively, for different system sizes. Both cumulants are seen to have non-monotonic behavior near the transition and go negative in this vicinity; this is another characteristic signature of a first order transition. We thus conclude that the layered to sublattice-ordered transition is first-order.

\section{\label{sec:layered}Correlations in layered phase}

\begin{figure}
\includegraphics[width=\columnwidth]{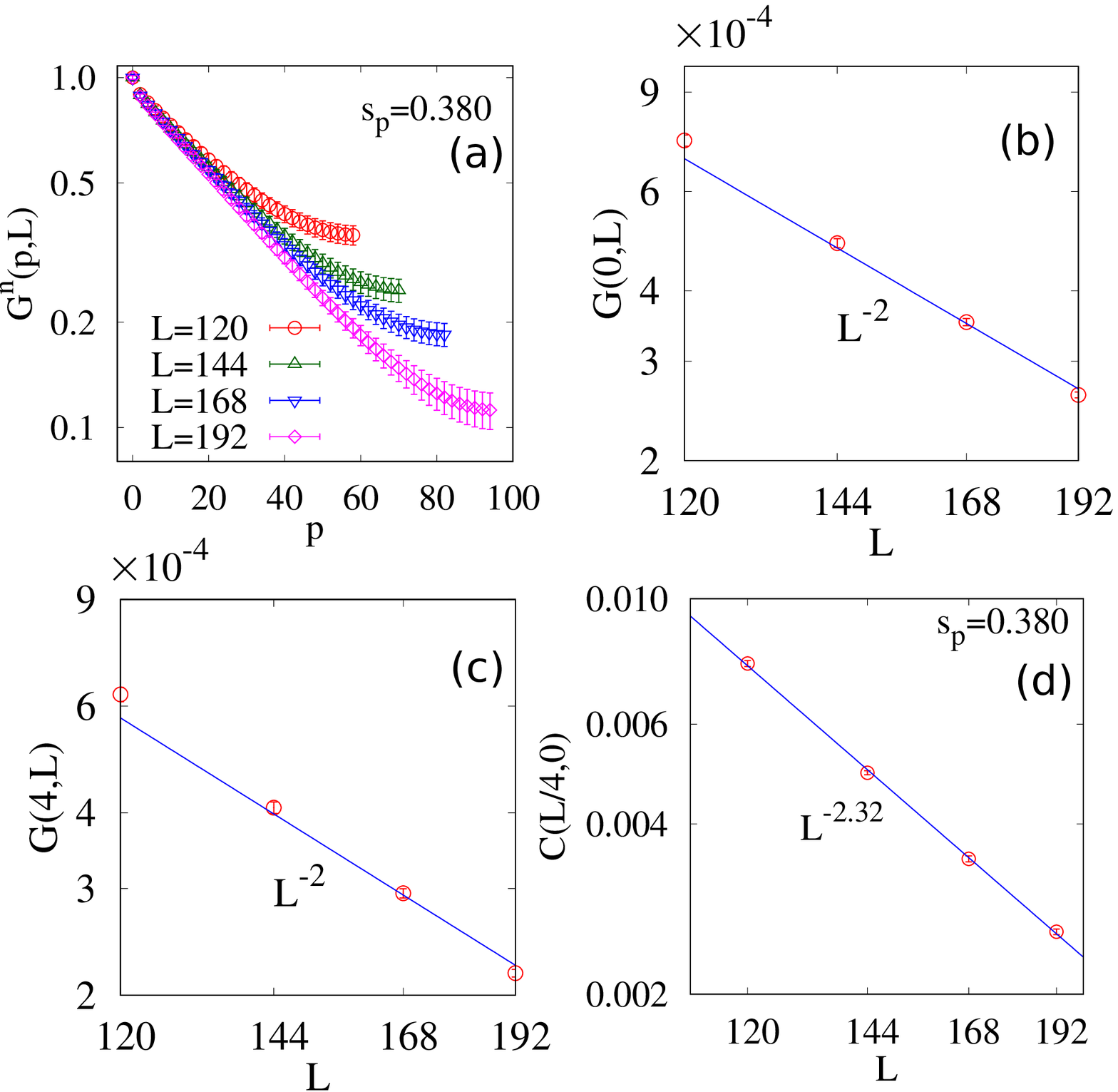}
\caption{(a) Variation of normalized layer correlation function $G_n(p, L)$ with interlayer separation $p$. (b) Variation of  unnormalized layer correlation function $G(p;L)$ for $p=0$, (c) Variation of  unnormalized layer correlation function $G(p;L)$ for $p=4$, and (d) in-plane two-point correlation function $C(r=L/4,p=0)$ as a function of $L$.}
\label{layered_correlation}
\end{figure}
In this section we characterize the correlations in the layered phase. To examine the intra-slab and inter-slab correlations, we define the in-plane columnar order parameter $(\ell_x(z) ,\ell_y(z))$ of a layer $z$ as 
\bea
\ell_x(z)=\frac{1}{L^2}\sum_{x,y=0}^{L-1}(-1)^x \phi(x,y,z),\nonumber\\
\ell_y(z)=\frac{1}{L^2}\sum_{x,y=0}^{L-1}(-1)^y \phi(x,y,z),
\eea
where $\phi(x,y,z)=1$, if the site is occupied by the head of a plate, and zero otherwise. The inter-slab correlation $G(p,L)$ for two slabs separated by a distance $p$ in the layering direction is defined as
\be
G(p,L)=\frac{1}{L} \sum_{z^\prime=0}^{L-1}[\ell_x(z^\prime) \ell_x(z^\prime+p) + \ell_y(z^\prime) \ell_y(z^\prime+p)],
\ee
The variation of the normalized correlation function $G^n(p,L)=G(p,L)/G(0,L)$ for the layered phase is shown in Fig.~\ref{layered_correlation}(a) for different systems sizes. It is clear that it decays exponentially with $p$. We conclude that the interaction between the slabs is weak and decays rapidly with inter-slab distance in the layered phase. The variation of $G(0,L)$ and $G(4,L)$ as a function of $L$ is shown in Fig.~\ref{layered_correlation}(b) and (c) respectively. For large $L$, we see that these approach the behavior
\be
G(p,L)\sim L^{-2} \;\;\; {\rm for} \;\;\; p=0,4 \; .
\ee

To understand the significance of this observation, we need to also study the correlation function of a local two-dimensional columnar order parameter field $\psi(r)$. To this end, we first note that each occupied slab, viewed along the layering axis, can be thought of as a two-dimensional system of hard squares and dimers; the plates with normals perpendicular to the layering axis play the role of dimers in this mapping, while plates with normal along the layering axis are viewed as hard squares. With this in hand, we employ the definition of $\psi$ used in Ref.~\cite{2015-rdd-prl-columnar} for a two-dimensional system of dimers and hard squares, and measure the connected two-point correlation function $C(r, p=0)$ of $\psi(r)$ within each occupied slab.
In Fig.~\ref{layered_correlation}(d), we display the $L$ dependence of $C(L/4,0)$ in the layered phase. As is clear from this figure, our data is consistent with $C(L/4,0) \sim 1/L^\eta$, with a $\eta > 2$ that depends on the plate fugacity.
\begin{figure}
\includegraphics[width=\columnwidth]{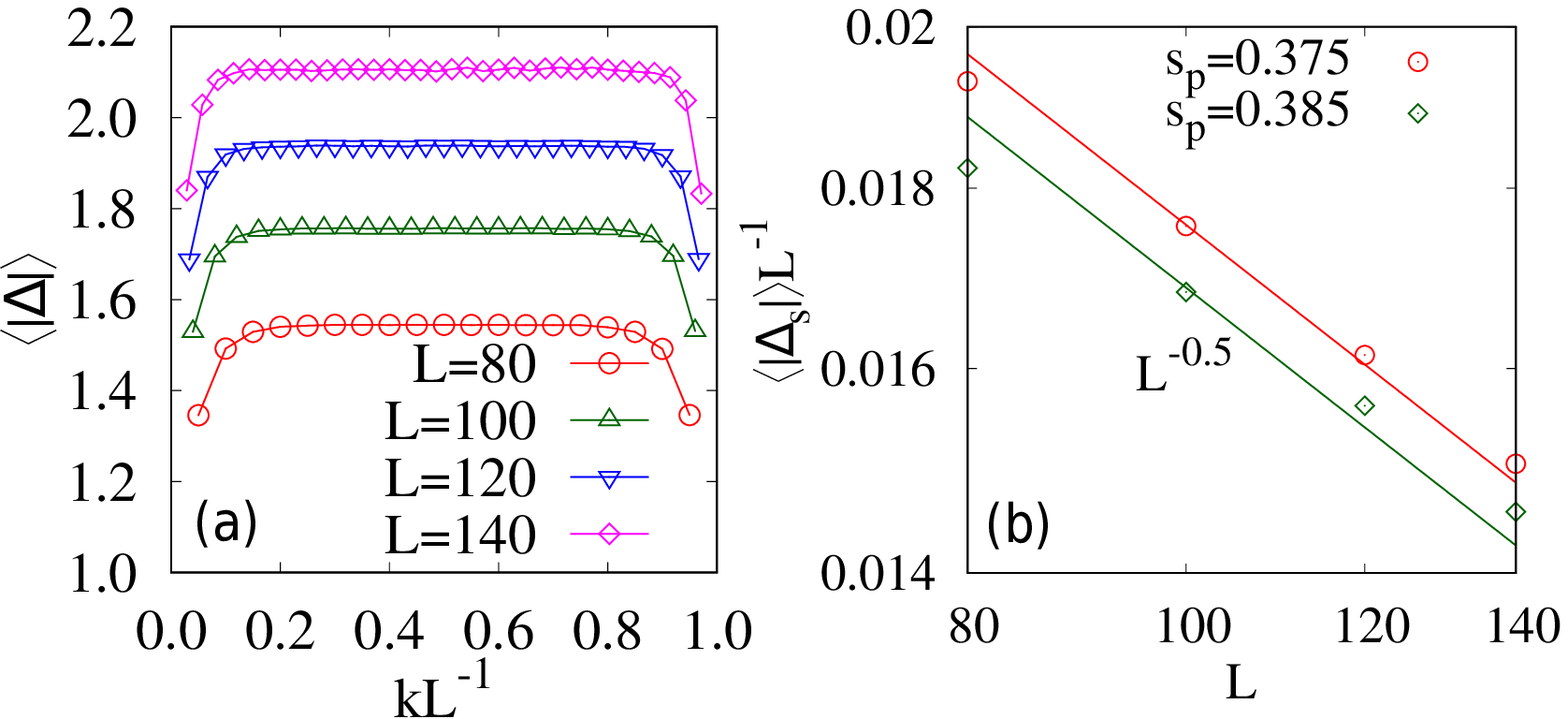}
\caption{(a) Variation of  the absolute value of total vacancy charge $|\Delta|$ in a $k\times L$ rectangular box on a occupied layer with $kL^{-1}$, for different $L$ in a layered phase with $s_p=0.375$. (b) Power-law scaling of the saturation charge $|\Delta_s|$ with $L$ for different $s_p$ in the layered phase.}
\label{fig:charge}
\end{figure}


This appearance of critical correlations in the layered phase at nonzero vacancy density is quite surprising at first sight. We understand it as a consequence of the constraints on the motion and relative positions of vacancies near full packing. Consider the time evolution of a system of hard plates at high density, with plates being only able to move to nearby empty spaces without causing any violations of the hard-core constraint; such a starting point is appropriate since the density of plates in the layered plates is quite high.  In such a system, vacancies can only move in pairs that can be thought of as dipoles. In the layered phase, the system splits into occupied slabs that are weakly coupled to each other, as is evident from the exponential falloff of the inter-slab correlations (Fig.~\ref{layered_correlation} a). Each occupied slab, viewed along the layering axis, is a system of hard squares and dimers on a two-dimensional square lattice. If individual vacancies move freely, such a two-dimensional system cannot support power-law order~\cite{2015-rdd-prl-columnar}. However, and this is key, dipolar defects do not destroy power-law order in this equivalent two-dimensional system. 

Motivated by this line of thought, we have monitored the total ``charge'' in a single layer (each such single layer forms the top layer or the bottom layer of an occupied slab). This total charge is defined as:
\be
|\Delta|=|\sum_{x,y \in {\rm strip of width k}}(-1)^{x+y}\delta_{\sigma,1} |,
\ee
where we have denoted the layering direction as $z$ (thus each layer is periodic in the $x$ and $y$ directions), the vacancy field $\sigma$ at a site is $1$ if a site is empty, and $0$ if it is touched by a plate, and the sum is taken over a strip that wraps around one periodic direction (perdendicular to the layering axis) of a $L \times L$ layer of our sample, and has finite width $k$ in the other periodic direction (again, perpendicular to the layering axis). The variation of average absolute charge as a function of $k/L$ for different $L$ and fixed $s_p=0.375$ is shown in Fig.~\ref{fig:charge}(a). The average absolute charge as a function of $k/L$ reaches a saturation value $|\Delta_s(L)|$ in the vicinity of $k/L = 0.5$. This function of $k/L$ is symmetric about $k/L=1/2$ because of the periodic boundary conditions obeyed by the layer. The $L$ dependence of $|\Delta_s(L)/L|$  is shown in Fig.~\ref{fig:charge}(b), and is seen to be consistent with a ``perimeter-law'' scaling. This perimeter-law scaling clearly admits a natural interpretation, namely that individual vacancies in any layer are bound into pairs, with each vacancy on a $A$ sublattice site paired with a nearby vacancy on the $B$ sublattice. Since charges are bound into dipolar pairs, $|\Delta_s(L)|$ naturally displays perimeter-law scaling with $L$.

However, the exponentially decaying correlations between slabs throws up another question. What prevents the power-law columnar order parameter of two adjacent occupied slabs from locking together? In the corresponding layered phase of fully-packed hard plates with anisotropic fugacityes, discussed in parallel work~\cite{Geetpaper}, the same question arises, and has an interesting answer: Namely, plates which straddle neighbouring occupied slabs occur in bound pairs, which can be viewed as quadrupolar defects. As a result, the corresponding coupling between slabs is irrelevant whenever the power-law exponent within the slab satisfies $\eta > 1/2$. 

With vacancies are allowed, as is the case here, there is a crucial difference: two adjacent occupied slabs can be coupled by single plates that straddle the two slabs. This is because a pair of vacancies can ``cut'' the string that binds two such plates into a quadrupolar defect in the fully-packed case~\cite{Geetpaper}.
Thus, in the present case, the question reduces to whether this dipole-dipole coupling between neighbouring occupied slabs is a relevant coupling.  From the scaling dimension of this dipole-dipole coupling, we see that this coupling is irrelevant whenever the transverse power-law exponent satisfies $\eta > 2$. Reassuringly, the measured value of $\eta$ throughout our layered phase does indeed satisfy this inequality. 
Thus it is the dipolar character of the defects within occupied slabs, and irrelevance of the dipolar couplings between slabs, that together lead to an a stable critical layered phase in this system. This critical layered phase is perhaps the most surprising aspect of the results presented here.

\section{\label{sec:discussion}Summary and discussion}
In this paper we  studied the phases and phase transitions (Fig.~\ref{schematic}) in a system of $2\times2\times1$ hard plates on the three dimensional cubic lattice using Monte Carlo simulations. Three types of plates are possible depending on their orientation, and our focus as been on the isotropic case with equal fugacity for all three types of plates. The system undergoes two phase transitions with increasing the density of particles: first, a continuous transition from disordered phase to layered phase that survives up to fairly high densities, and second, a first-order transition from the layered phase to a sublattice-ordered phase that is stable at~\cite{Geetpaper} and near full-packing. 

In the sublattice-ordered phase, the system displays two-fold breaking of translational symmetry along all three cartesian axis. Each type of plate has columnar order, and breaks translation symmetry in the two directions perpendicular to its axis [see Fig.~\ref{time_pf_sub}]. In the  layered phase, the density of one type of plate is lower relative to the other two, and there is two-fold translation symmetry breaking along one spontaneously chosen cartesian axis, with occupied slabs (with a higher density of plates contained entirely within them) separated from each other by one lattice spacing as one moves along this layering axis. 
Remarkably, correlations within an occupied slab decay as an oscillatory power law, with wavevector corresponding to power-law columnar order within the slab. On the other hand,  the correlations between different occupied slabs decrease exponentially with the separation between them. 

As mentioned earlier in Sec~\ref{sec:layered} (see also Ref.~\cite{Geetpaper}), this vacancy-driven physics of hard plates on the cubic lattice is particularly interesting from a vantage point that uses the fully-packed system as a reference and views the vacancies as defects introduced into a fully-packed configuration. This is best appreciated by contrasting the constraints on the position and mobility of individual vacancies in this system with the corresponding constraints (or lack thereof) in systems of $k$-mers ($k>2$)~\cite{2007-gd-epl-on,2013-krds-pre-nematic,dhar2021entropy} or dimers ($k=2$)~\cite{1961-k-physica-statistics,1961-tf-pm-dimer,2003-hkms-prl-coulomb,2017-naq-arxiv-polyomino}. 

Consider removing a single dimer from a fully-packed dimer model on the bipartite square or cubic lattice. This introduces two vacancies, one on the A sublattice and the other on the $B$ sublattice of the bipartite lattice. As the dimers move around while obeying the hard-core constraint on their positions, the two vacancies can separate from each other and move individually via hops to next-nearest-neighbor sites. In other words, the only constraint on them is that the two vacancies must occupy opposite sublattices. Turning to long rigid rods of length $k$ with $k>2$, the situation is not very different: Consider the $k$ vacancies, created by the removal of a single rod from the fully-packed system. Apart from some constraints on the sublattices of sites that can be simultaneously occupied by these $k$ vacancies, these vacancies can move around and separate from one another.
This should be contrasted with the constraints faced by the four vacancies that are created when a single hard plate is removed from the fully-packed system on the cubic lattice. These vacancies are only free to move as two nearest neighbor pairs, and that too only in directions perpendicular to the pairing axis. This is a key distinction between the present problem and systems of long rods. Indeed, the problem studied here has a stable sublattice-ordered phase at densities close to full-packing, while in the case of long rods, the sublattice-ordered phase  is unstable close to full-packing because of a sliding instability.

In bipartite dimer models, each dimer can be thought of as a dipole, and the fully-packed limit is understood in terms of a coarse-grained height action that describes the potential field in a system of fluctuating dipoles. This provides a natural description of the Coulomb correlations of bipartite dimer models~\cite{Henley_coulombphasereview,Alet_2006pre_interactingdimers,Papa_Fradkin_interactingdimers,2015-rdd-prl-columnar,Huse_etal_3ddimers,Desai_Pujari_Damle_bilayerdimers}. Isolated vacancies correspond to charged monopoles in this description. Any nonzero density of vacancies then corresponds to a nonzero density of free charges, which introduces a finite correlation length and destroys the Coulomb liquid phase. 
Although less is known, the effect of a small density of vacancies on fully-packed k-mers is expected to be quite similar in two dimensions, since the full-packing limit again admits a multi-component height description in the two-dimensional case, and isolated vacancies now correspond to vector charges~\cite{2007-gd-epl-on,2013-krds-pre-nematic} within this description. In contrast, since vacancies in our fully-packed plate system can only move in pairs, there is no ``free charge'' associated with them. Instead, pairs of vacancies in a layer are more appropriately thought of as dipolar defects in the coarse-grained effective field theory~\cite{Geetpaper} for a layer.
Thus, our results, particularly the transition to the spontaneously layered phase and the critical correlations of the occupied slabs, can be viewed as being a direct consequence of this restricted motion of vacancy defects in the hard plate lattice gas; this point of view is particularly appropriate since the transition to the critical layered phase occurs at a very small vacancy density of $\rho_{\rm vac}^{\rm crit} = 0.026$.


Finally, we note that although numerous analytical~\cite{1973-a-prl-phase,1974-s-pra-ordered,2018-dgj-araiv-plate}, experimental~\cite{1980-ys-prl-observation,2004-l-nature-missing} and computer simulation~\cite{2000-bz-jcp-thermotropic,2008-bmorz-jpcm-computer,2018-dtdrd-prl-hard} studies indicate the presence of biaxial nematic phase (in which the system exhibits orientational order along all three internal axis of the particle) in systems of anisotropic plate-like objects in three dimensions, there has been some debate regarding the existence of this phase. In this paper, for the particular case of $2 \times 2 \times 1$ hard plates on the cubic lattice, we have not found any biaxial nematic phase. It would be very interesting to study the existence and stability of such a biaxial nematic phase for other lattice models of plates which cover more than one elementary face of the cubic lattice. A system of rectangular plates with different aspect ratio having a hard core and/or attractive interaction would also be a promising candidate for future study. 

\section*{Acknowledgments}
We thank  K. Ramola and N. Vigneshwar for helpful discussions.
The simulations were carried out on the high performance computing machines Nandadevi at the Institute of Mathematical Sciences, and the computational facilities provided by the University of Warwick Scientific Computing Research Technology Platform. Some of this work contributed to the Ph.D thesis of DM submitted to the Homi Bhabha National Institute (HBNI).  GR was supported by the TQM unit of Okinawa Institute of Science and Technology during the final stages of this work. KD was supported at the
TIFR by DAE, India and in part by a J.C. Bose Fellowship
(JCB/2020/000047) of SERB, DST India, and by
the Infosys-Chandrasekharan Random Geometry Center
(TIFR). D.D.’s work was partially supported by Grant No. DST-SR-
S2/JCB-24/2005 of the Government of India, and partially by a Senior Scientist Fellowship by  the National Academy of Sciences of India.

{\em Author contributions} DM performed the computations with assistance from GR. KD, RR, and DD conceived and directed this work, and finalized the manuscript using detailed inputs from DM.

\bibliography{references}
\end{document}